\documentclass[conference]{IEEEtran}

\usepackage[dvipsnames]{xcolor}

\usepackage{mathtools}
\usepackage{booktabs}
\usepackage{pgfplots}
\pgfplotsset{compat = newest}

\usepackage{amsmath}
\usepackage{amsfonts}
\usepackage{float}

\usepackage{listings}

\definecolor{gray}{rgb}{0.5,0.5,0.5}
\definecolor{mauve}{rgb}{0.58,0,0.82}
\definecolor{lightgrey}{rgb}{0.9,0.9,0.9}
\definecolor{darkgreen}{rgb}{0,0.6,0}

\usepackage{tikz}
\usetikzlibrary{positioning,fit,calc}

\usepackage{pgfplots}
\usepgfplotslibrary{groupplots}

\usepackage[hyphens]{url}
\usepackage[hidelinks]{hyperref}
\usepackage[hyphenbreaks]{breakurl}

\DeclareMathOperator{\fl}{\mathrm{fl}}

\newcommand{\MS}[1]{\mathbb{#1}}

\title{MATLAB Simulator of Level-Index Arithmetic}

\begin{document}

\author{
\IEEEauthorblockN{
Mantas Mikaitis}
\IEEEauthorblockA{
  School of Computing, University of Leeds, Leeds, United Kingdom
}
}

\maketitle

\begin{abstract}
  Level-index arithmetic appeared in the 1980s.
  One of its principal purposes is to abolish the issues caused by underflows and overflows in floating point.
  However, level-index arithmetic does not expand the set of numbers but spaces out the numbers of large magnitude even more than floating-point representations to move the infinities further away from zero: gaps between numbers on both ends of the range become very large.
  We revisit level index by presenting a custom precision simulator in MATLAB.
  This toolbox is useful for exploring performance of level-index arithmetic in research projects, such as using 8-bit and 16-bit representations in machine learning algorithms where narrow bit-width is desired but overflow/underflow of floating-point representations causes difficulties.
\end{abstract}


\IEEEdisplaynontitleabstractindextext

\section{Introduction}

In a level-index (LI) arithmetic of Clenshaw~and~Olver~\cite{clol84} a positive number $x \in \MS{R}$ is represented with $l \in \MS{N}$ (a level) and $f \in [0,1)$ (an index) as
\begin{equation}
  x = e^{e^{.^{.^{.^{e^{f}}}}}}.
  \label{eq:level-index-x}
\end{equation}
Here
\begin{equation}
  f = \ln(\ln(\cdots \ln(x) \cdots)).
  \label{eq:level-index-f}
\end{equation}
The exponentiation or the logarithm are taken $l$ times.
Numbers $x<1$ could be represented by taking $l=0$ and $f=x$, but more of them can be represented by the symmetric level-index (SLI) system by Clenshaw~and~Turner~\cite{cltu88} which adds a reciprocal sign to \eqref{eq:level-index-x} used for $x<1$.
The term ``symmetric'' presumably refers to the numbers of values represented in the ranges $x \in (0, 1)$ and $x \in (1, +\infty)$ being the same; the ordinary LI arithmetic that does not use the reciprocal sign is also symmetric, but with respect to zero when taking into account the negative axis.

Formally, a nonzero real number $x$ in the SLI systems is represented by a number $\zeta=l+f$ and the following relations~\cite{cltu88}:
\begin{equation}
  x=s(x)\phi (\zeta)^{r(x)},
  \label{eq:level-index-main}
\end{equation}
where $s(x) = \pm 1$ is the sign of $x$, $r(x) = \pm 1$ is the reciprocal sign defined by
\begin{align}
  r(x)=
  \begin{cases}
    +1, & \text{if } |x| \geq 1, \\
    -1, & \text{if } |x| < 1,
  \end{cases}
\end{align}
and
\begin{align}
  \phi(\zeta)=
  \begin{cases}
    \zeta, & \text{if } 0 \leq \zeta < 1, \\
    e^{\phi(\zeta-1)}, & \text{if } \zeta \geq 1.
  \end{cases}
  \label{eq:level-index-phi}
\end{align}
Here \eqref{eq:level-index-phi} computes \eqref{eq:level-index-x} given a LI number.
Note that \eqref{eq:level-index-phi} produces $\lfloor \zeta \rfloor = l$ exponentials, with the final exponent $\zeta - \lfloor \zeta \rfloor = f$ as required by the definition of the LI systems.

To construct $\zeta$ Clenshaw~and~Olver~\cite{clol84} propose
\begin{align}
  \Psi(x)=
  \begin{cases}
    x, & \text{if } 0 \leq x < 1, \\
    1 + \Psi(\ln(x)), & \text{if } x \geq 1,
  \end{cases}
  \label{eq:level-index-psi}
\end{align}
which is similar to \eqref{eq:level-index-f} except that the level is also included with $1$ being added on every recursive step.

Note that precision $p$ does not come in anywhere in this definition, unlike the floating-point representation that usually contains $p$.
Of course, $p$ plays a role in implementing the quantisation of the index $f$.


\subsection{Previous results}

Turner~\cite{turn89} demonstrates a Pascal software package that simulates a SLI format with 3 level bits and 27 index bits.

Lozier~and~Olver~\cite{olve87,lool90} show that LI system is \emph{closed} which means that, unlike in floating point, it is impossible to produce numbers that lie outside the representable range with the basic operations, except division by zero.
The authors~\cite[Sec.~3]{olve87} also explain that levels beyond 6 bits will not be entered in practice by addition, subtraction, multiplication and division, and therefore that 3 bits are enough for the level.
Furthermore, Olver~\cite[Sec.~4]{olve87} writes that LI systems are free from ``wobbling precision'', a feature of floating point whereby a real number $x$ rounded to a floating-point system with precision $p$ is $\fl(x)=x(1+\delta)$ and the error $\delta$ can be anywhere between $-2^{-p}$ and $2^{-p}$.
Olver also mentions that LI is more precise than floating-point for $x<2^{11}$ in 32 bits and for $x<2^{44}$ in 64 bits, but less precise beyond $x > 2^{18}$ and $x > 2^{70}$ for 32- and 64-bit representations, respectively.

Demmel~\cite{demm87} argues that LI and other similar arithmetics such as the one by Iri and Matsui~\cite{mair81} that aim to remove the possibility of overflow, overall do not result in improvements since more care is needed when computing with very big highly inaccurate quantities.
This is in contrast with floating point that returns infinities allowing to detect overflows.

Shen~and~Turner~\cite{shtu06} explore a hybrid floating point and LI arithmetic.
They propose to do most computations in standard binary64~\cite{ieee19} arithmetic, but switch to LI once certain bounds are reached on the input arguments to the four basic arithmetic operations.

Kwak~and~Swartzlander~\cite{kwsw98} propose a hardware implementation of LI arithmetic and demonstrate area reduction with a minor increase in the timing of the circuit, compared with the previous approach by Olver~and~Turner~\cite{oltu87}.

\section{The encoding of level-index numbers}

We need to encode the LI numbers of \eqref{eq:level-index-main} in a limited precision word length for use on the digital computers.
A sign $s(x)$ can be one bit, the reciprocal sign $r(x)$ can also be one bit, the level $l$, the authors recommend, does not have to be more than 3 bits to ``virtually abolish overflow from everyday work'' \cite{clol84}, and the index $f$ can be as precise as possible and represented in fixed point.
We will refer to a sli encoding with a $k$-bit level and a $p$-bit index as sli-$k$.$p$.

\begin{figure}
  \centering
  \footnotesize
  \begin{tikzpicture}
    \node (s) [rectangle,draw, minimum height = 1cm, minimum width = 0.5cm] {$s$};
    \node (r) [rectangle,draw, minimum height = 1cm, minimum width = 0.5cm, right=-\the\pgflinewidth of s.east] {$r$};
    \node (l) [rectangle,draw, minimum height = 1cm, minimum width = 1.5cm, right=-\the\pgflinewidth of r.east] {$l$};
    \node (f) [rectangle,draw, minimum height = 1cm, minimum width = 5.5cm, right=-\the\pgflinewidth of l.east] {$f$};
    \draw[decorate,decoration={brace}, rotate=180] ($(s)+(-0.2,0.6)$) -- ($(s)+(0.2,0.6)$)
    node[below=2pt, pos=0.5]{$1$};
    \draw[decorate,decoration={brace}, rotate=180] ($(r)+(-0.2,0.6)$) -- ($(r)+(0.2,0.6)$)
    node[below=2pt, pos=0.5]{$1$};
    \draw[decorate,decoration={brace}, rotate=180] ($(l)+(-0.6,0.6)$) -- ($(l)+(0.6,0.6)$)
    node[below=2pt, pos=0.5]{$3$};
    \draw[decorate,decoration={brace}, rotate=180] ($(f)+(-2.7,0.6)$) -- ($(f)+(2.7,0.6)$)
    node[below=2pt, pos=0.5]{$11$ bits};
  \end{tikzpicture}
  \caption{Layout of a possible 16-bit SLI encoding: sli-3.11.}
  \label{fig:level-index-encoding}
\end{figure}
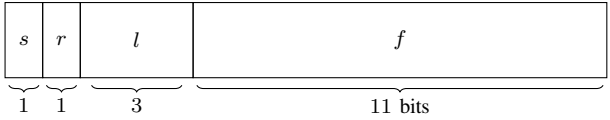

Figure~\ref{fig:level-index-encoding} shows an encoding of a 16-bit binary SLI representation with a 3-bit level and an 11-bit index.
We have placed the reciprocal sign $r(x)$ to the left of the level.
This way the sequence of representable numbers starts from the smallest number, representing $r(x)=-1$ by setting the reciprocal bit to zero.
Then, when the encoding bit pattern is incremented by 1, it transitions through the levels and eventually $r(x)=1$ is set when the representation for number $1$ is reached.

\section{Small level-index systems}

Here we compare an unsigned 5-bit SLI representation with an unsigned 5-bit binary floating-point (``toy'') system used for demonstration by Higham~\cite{high21}.
In the interest of saving space we don't include the negative axes---the representations are symmetrical with respect to zero.
In floating point, numbers are represented with $\pm \beta^{e-p+1}\times m$.
Here $\beta$ is a base, $p$ is precision, and $e_{min} \leq e \leq e_{max}$ is the exponent.
The exponent is usually encoded with a bias: $E=e+e_{max}$.
In IEEE 754 \cite{ieee19} $e_{min}=1-e_{max}$.
The significand $m$ satisfies $0 \leq m \leq \beta^p-1$, but the normalized nonzero numbers are assumed to have $m \geq \beta^{p-1}$ whilst the subnormal values have $m \leq \beta^{p-1}-1$ and a fixed exponent $e=e_{min}$.
Stored significand $M$ omits the most significant bit of $m$.
We will assume the IEEE 754 floating-point encoding, including representing infinities, subnormals and not-a-number (NaNs).

For the 5-bit SLI representation, we consider 1 and 2 bits for the level.
Since the encoding for level zero is not required, we use the level encoding $00$ for representing level $1$ and therefore have levels $1$ to $4$.
Similarly for $1$-bit level SLI representations: $0$ encodes level $1$ and $1$ encodes level $2$.
The authors of LI mention that it could instead be used for representing special values \cite{cltu88}, perhaps values equivalent to NaNs or infinities in floating point.
For the reciprocal bit sign, we use $0$ for $r(x)=-1$ and $1$ for $r(x)=1$ and store it on the left of the level bits in order to have small values represented by the lower half of the set of representable binary patterns.

\begin{figure}
  \centering
  \footnotesize
  \begin{tabular}{c}
  \begin{tikzpicture}
    \node (e) [rectangle,draw, minimum height = 1cm, minimum width = 1.5cm] {$E$};
    \node (m) [rectangle,draw, minimum height = 1cm, minimum width = 1cm, right=-\the\pgflinewidth of e.east] {$M$};
    \draw[decorate,decoration={brace}, rotate=180] ($(e)+(-0.6,0.6)$) -- ($(e)+(0.6,0.6)$)
    node[below=2pt, pos=0.5]{$3$};
    \draw[decorate,decoration={brace}, rotate=180] ($(m)+(-0.4,0.6)$) -- ($(m)+(0.4,0.6)$)
    node[below=2pt, pos=0.5]{$2$ bits};
  \end{tikzpicture} \\
  \end{tabular}

  \begin{tabular}{cc}
  \begin{tikzpicture}
    \node (r) [rectangle,draw, minimum height = 1cm, minimum width = 0.5cm] {$r$};
    \node (l) [rectangle,draw, minimum height = 1cm, minimum width = 0.5cm, right=-\the\pgflinewidth of r.east] {$l$};
    \node (f) [rectangle,draw, minimum height = 1cm, minimum width = 2cm, right=-\the\pgflinewidth of l.east] {$f$};
    \draw[decorate,decoration={brace}, rotate=180] ($(r)+(-0.2,0.6)$) -- ($(r)+(0.2,0.6)$)
    node[below=2pt, pos=0.5]{$1$};
    \draw[decorate,decoration={brace}, rotate=180] ($(l)+(-0.2,0.6)$) -- ($(l)+(0.2,0.6)$)
    node[below=2pt, pos=0.5]{$1$};
    \draw[decorate,decoration={brace}, rotate=180] ($(f)+(-0.8,0.6)$) -- ($(f)+(0.8,0.6)$)
    node[below=2pt, pos=0.5]{$3$ bits};
  \end{tikzpicture} &
  \begin{tikzpicture}
    \node (r) [rectangle,draw, minimum height = 1cm, minimum width = 0.5cm] {$r$};
    \node (l) [rectangle,draw, minimum height = 1cm, minimum width = 1cm, right=-\the\pgflinewidth of r.east] {$l$};
    \node (f) [rectangle,draw, minimum height = 1cm, minimum width = 1.5cm, right=-\the\pgflinewidth of l.east] {$f$};
    \draw[decorate,decoration={brace}, rotate=180] ($(r)+(-0.2,0.6)$) -- ($(r)+(0.2,0.6)$)
    node[below=2pt, pos=0.5]{$1$};
    \draw[decorate,decoration={brace}, rotate=180] ($(l)+(-0.4,0.6)$) -- ($(l)+(0.4,0.6)$)
    node[below=2pt, pos=0.5]{$2$};
    \draw[decorate,decoration={brace}, rotate=180] ($(f)+(-0.6,0.6)$) -- ($(f)+(0.6,0.6)$)
    node[below=2pt, pos=0.5]{$2$ bits};
  \end{tikzpicture}
  \end{tabular}
  \caption{Layout of an unsigned 5-bit toy floating-point system \cite{high21} (top) and two unsigned SLI systems: sli-1.3 (bottom left) and sli-2.2 (bottom right).}
  \label{fig:toy-fp-system}
\end{figure}
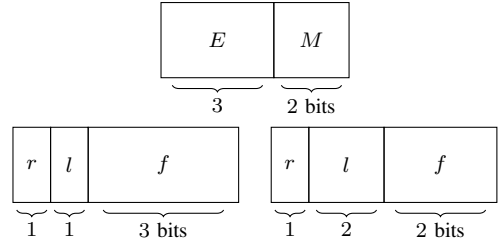

Figure~\ref{fig:toy-fp-system} shows the layouts of 5-bit floating-point representation and two different 5-bit SLI representations.
Table~\ref{table:toy-system-values} lists the 32 possible values representable by the three systems.
The following observations can be made from this table.

\begin{itemize}
\item SLI systems have two representations for $1$ and no representation for $0$. One of the bit patterns for $1$ could be used for representing zero. We have used all zeros to represent the zero.
\item SLI systems represent decreasing numbers with the increasing binary patterns from $00000$ to $01111$, which happens because of the reciprocal rule for representing small values below $1$. This could be changed by inverting the index bits on conversion to and from, if $r(x)=0$. Then $00000$ could be used to represent the zero, with the following number $00001$, when inverted, representing the smallest representable nonzero number. We did not implement this in v0.1 of the toolbox to keep the encoding closer to the definition~\eqref{eq:level-index-main}.
\item The SLI system with the 1-bit level does not offer a wider dynamic range than the 5-bit floating-point system.
\item The SLI system with the 2-bit level offers a wider dynamic range than the binary64~\cite{ieee19} representation.
\end{itemize}

\begin{table}[ht!]
  \centering
  \caption{All quantities encoded in the toy 5-bit floating-point and SLI systems of Figure~\ref{fig:toy-fp-system}.}
  \begin{tabular}{llrr}
    \toprule
     & FP & sli-1.3 & sli-2.2 \\
    \midrule
    $00000$ & $0$ & $(e^{0})^{-1}=1$ & $(e^{0})^{-1}=1$ \\
    $00001$ & $0.0625$ & $(e^{0.125})^{-1}\approx 0.8825$ & $(e^{0.25})^{-1}\approx 0.7788$ \\
    $00010$ & $0.125$ & $(e^{0.25})^{-1}\approx 0.7788$ & $(e^{0.5})^{-1}\approx 0.6065$ \\
    $00011$ & $0.1875$ & $\sim 0.6873$ & $\sim 0.4724$ \\
    $00100$ & $0.25$ & $\sim 0.6065$ & $(e^{e^{0}})^{-1}\approx 0.3679$ \\
    $00101$ & $0.3125$ & $\sim 0.5353$ & $\sim 0.2769$ \\
    $00110$ & $0.375$ & $\sim 0.4724$ & $\sim 0.1923$ \\
    $00111$ & $0.4375$ & $\sim 0.4169$ & $\sim 0.1204$ \\
    $01000$ & $0.5$ & $(e^{e^{0}})^{-1}\approx 0.3679$ & $(e^{e^{e^{0}}})^{-1}\approx 0.06599$ \\
    $01001$ & $0.625$ & $(e^{e^{0.125}})^{-1}\approx 0.322$ &  $\sim 0.02702$ \\
    $01010$ & $0.75$ & $(e^{e^{0.25}})^{-1}\approx 0.2769$ &  $\sim 0.0055$ \\
    $01011$ & $0.875$ & $\sim 0.2334$ &  $\sim 2.4 \times 10^{-4}$ \\
    $01100$ & $1$ & $\sim 0.1923$ & $(e^{e^{e^{e^{0}}}})^{-1}\approx 2.6 \times 10^{-7}$ \\
    $01101$ & $1.25$ & $\sim 0.1544$ & $\sim 8.4 \times 10^{-17}$ \\
    $01110$ & $1.5$ & $\sim 0.1204$ & $\sim 1.7 \times 10^{-79}$\\
    $01111$ & $1.75$ & $\sim 0.0908$ & $\sim 10^{-1758}$\\
    $10000$ & $2$ & $(e^{0})^{1}=1$ & $(e^{0})^{1}=1$ \\
    $10001$ & $2.5$ & $(e^{0.125})^{1}\approx 1.1331$ & $(e^{0.25})^{1}\approx 1.284$ \\
    $10010$ & $3$ & $(e^{0.25})^{1}\approx 1.284$ &  $(e^{0.5})^{1}\approx 1.6487$ \\
    $10011$ & $3.5$ & $\sim 1.455$ & $\sim 2.117$ \\
    $10100$ & $4$ & $\sim 1.6487$ & $(e^{e^{0}})^{1}\approx 2.7183$ \\
    $10101$ & $5$ & $\sim 1.8682$ & $\sim 3.6111$ \\
    $10110$ & $6$ & $\sim 2.117$ & $\sim 5.2003$ \\
    $10111$ & $7$ & $\sim 2.3989$ & $\sim 8.3062$ \\
    $11000$ & $8$ & $(e^{e^{0}})^{1}\approx 2.7183$ & $(e^{e^{e^{0}}})^{1}\approx 15.1533$ \\
    $11001$ & $10$ & $(e^{e^{0.125}})^{1}\approx 3.1054$ & $\sim 37.0085$ \\
    $11010$ & $12$ & $(e^{e^{0.25}})^{1}\approx 3.6111$ & $\sim 181.3313$ \\
    $11011$ & $14$ & $\sim 4.2844$ & $\sim 4048.8237$ \\
    $11100$ & $+\infty$ &  $\sim 5.2$ & $(e^{e^{e^{e^{0}}}})^{1}\approx 3.8 \times 10^{6}$ \\
    $11101$ & NaN &  $\sim 6.4769$ & $\sim 1.18 \times 10^{16}$\\
    $11110$ & NaN &  $\sim 8.306$ & $\sim 5.6387 \times 10^{78}$ \\
    $11111$ & NaN & $\sim 11.0108$ & $\sim 10^{1758}$ \\
    \bottomrule
  \end{tabular}
  \label{table:toy-system-values}
\end{table}

\section{Arithmetic with level-index numbers}

The algorithms for LI arithmetic are shown by Clenshaw~and~Olver~\cite{clol87} while the modifications required for the SLI systems are investigated by Clenshaw~and~Turner~\cite{cltu88}.
We provide key highlights to demonstrate what is involved in implementing LI arithmetic; readers should refer to Clenshaw, Olver, and Turner for complete algorithms.

Take $X=l+f$, $Y=m+g$, $Z=n+h$, $X \geq Y \geq 0$, the LI numbers with corresponding levels $l$, $m$ and $n$, indices $f$, $g$ and $h$, and $\phi(X)\pm\phi(Y)=\phi(Z)$.
In the standard LI arithmetic, addition and subtraction operations  require three sequences $a_j=1/\phi(X-j)$, $b_j=\phi(Y-j)/\phi(X-j)$, and $c_j=\phi(Z-j)/\phi(X-j)$.
Sequences terminate as soon as $c_j < a_j$, and additional calculations on $c_j$ provide the level and index values of the final result~\cite{clol87} (rounding or chopping to required precision).
These sequences are short because for $a_j$ and $c_j$ $j$ goes up to the level of $X$ while for $b_j$ up to the level of $Y$.
Clenshaw~and~Olver~\cite{clol87} provide

\begin{equation*}
  a_{l-1}=e^{-f}, \; a_{j-1}=e^{-1/a_j},
\end{equation*}
\begin{equation*}
  b_{m-1}=a_{m-1}e^g, \; b_{j-1}=e^{-(1-b_j)/a_j} \; (\text{if } m \geq 1), \; \text{and}
\end{equation*}
\begin{equation*}
  c_0=1-b_0, \; c_j=1+a_j\ln(c_{j-1}).
\end{equation*}
If $m=0$, we compute $b_0=a_0g$ instead of the expression above. For addition $c_0=1+b_0$.
The sequence $c_j$ is stopped when $c_j<a_j$, at which point $n=j$ and $h=c_j/a_j$.
If $c_j\geq a_j$ for $j = 0, \dots, l-1$, then $n=l$ and $h=f+\ln(c_{l-1})$ \cite{clol87}.

Multiplication and division are straightforward~\cite{clol87}: with extra manipulations of arguments we turn the operations into addition or subtraction and therefore reuse the sequences above.
For example, if $m > 0$ then $n > 0$ and $\phi(X)\phi(Y)=\phi(Z)=e^{\phi(X-1)}e^{\phi(Y-1)}=e^{\phi(Z-1)}$ allows to write $\phi(X-1)+\phi(Y-1)=\phi(Z-1)$.
We can then do the addition and increase the level of $Z-1$ by one.
Further details are in \cite{clol87}.

Arithmetic for SLI systems requires a few modifications since there is no level zero and the reciprocal sign has to be taken into account.
These modifications are described by Clenshaw~and~Turner~\cite{cltu88}.

\section{MATLAB symmetric level index \texttt{sli.m}}

We have implemented a simulator for the SLI arithmetic \cite{cltu88} in MATLAB.
Version 0.1 is available on GitHub\footnote{ \url{https://github.com/north-numerical-computing/level-index-simulator.git}}.
The file \texttt{sli.m} defines a \texttt{sli} object, with the following properties.
\begin{itemize}
\item \texttt{level\_bits}: number of bits assigned to the level ($p_l$). By default it is set to $2$.
\item \texttt{index\_bits}: number of bits assigned to the index ($p_i$). By default it is set to $12$.
\item \texttt{sign}: sign bit.  0 for $s(x)=1$ and $1$ for $s(x)=-1$.
\item \texttt{reciprocal}: 1 for $r(x)=1$ and $0$ for $r(x)=-1$.
\item \texttt{level}: level, stored as binary64, limited to $[1, 2^{p_l}]$. Since it is a positive integer it could be stored as a 64-bit integer.
\item \texttt{index}: index, stored as binary64, rounded to a fixed-point representation with machine epsilon $\varepsilon=2^{-p_i}$ using MATLAB's \texttt{round()} (round-to-nearest ties-to-away, a default rounding mode). This value could be stored as a 64-bit integer in fixed-point representation.
\item \texttt{value}: a binary64 image of the stored LI number, constructed using \eqref{eq:level-index-phi}.
\end{itemize}

Below is an example use of \texttt{sli} in MATLAB.
\begin{lstlisting}
>> x=sli
[...]
>> x=x.set_val(pi)
x = 
  sli with properties:
    level_bits: 2
    index_bits: 12
          sign: 0
    reciprocal: 1
         level: 2
         index: 0.135253906250000
         value: 3.141899100868418
>> x*x
ans = 
  sli with properties:
    level_bits: 2
    index_bits: 12
          sign: 0
    reciprocal: 1
         level: 2
         index: 0.828369140625000
         value: 9.870807937639510
\end{lstlisting}

There are two ways to define a \texttt{sli} object: by specifying a binary64 quantity or by explicitly specifying the level and index values.
The first method uses \eqref{eq:level-index-psi} to convert a binary64 value to a LI value, with rounding to nearest for fitting the index into the specified number of bits.
See the example code within the repository for more detail.

\section{Experiments}

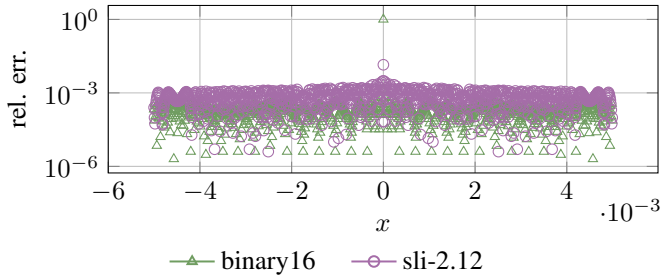
\begin{figure}[ht!]
  \begin{center}
    \begin{tikzpicture}
      \begin{groupplot}[
        group style={
          group size=1 by 1,
        },
        width=3.5in,
        height=1.5in,
        grid=major,
        ]
        
        \nextgroupplot[
        ymode=log,
        ylabel={rel. err.},
        xlabel = {$x$},
        every x tick scale label/.style={
          at={(axis description cs:0.95,-0.1)},
          anchor=north
        }
        ]
        
        \addplot[color=OliveGreen!70, mark=triangle, only marks] table [x=x, y=binary16] {data/representation_binary16.dat};
        \addplot[color=Fuchsia!70, mark=o, only marks] table [x=x, y=level-index] {data/representation_binary16.dat};
        
      \end{groupplot}
    \end{tikzpicture}
    
    \begin{tikzpicture}[trim axis left, trim axis right]
      \begin{axis}[
        title = {},
        legend columns=2,
        scale only axis,
        width=1mm,
        hide axis,
        /tikz/every even column/.append style={column sep=0.4cm},
        legend style={at={(0,0)},anchor=center,draw=none,
          legend cell align={left},cells={line width=0.75pt}},
        legend image post style={sharp plot},
        legend cell align={left},
        ]
        \addplot [mark=triangle, OliveGreen!70] (0,0);
        \addplot [mark=o, Fuchsia!70] (0,0);
        \legend{binary16, sli-2.12};
      \end{axis}
    \end{tikzpicture}
  \end{center}
  \caption{Relative accuracy of binary16 and a 16-bit level-index representation compared with binary64.}
  \label{fig:representation-accuracy-binary16}
\end{figure}

\begin{figure}[ht!]
  \begin{center}
    \begin{tikzpicture}
      \begin{groupplot}[
        group style={
          group size=1 by 1,
        },
        width=3.5in,
        height=1.5in,
        grid=major,
        ]
        
        \nextgroupplot[
        ymode=log,
        ylabel={rel. err.},
        xlabel = {$x$},
        every x tick scale label/.style={
          at={(axis description cs:0.95,-0.1)},
          anchor=north
        }
        ]
        
        \addplot[color=OliveGreen!70, mark=triangle, only marks] table [x=x, y=bfloat16] {data/representation_bfloat16.dat};
        \addplot[color=Fuchsia!70, mark=o, only marks] table [x=x, y=level-index] {data/representation_bfloat16.dat};
        
      \end{groupplot}
    \end{tikzpicture}
    
    \begin{tikzpicture}[trim axis left, trim axis right]
      \begin{axis}[
        title = {},
        legend columns=2,
        scale only axis,
        width=1mm,
        hide axis,
        /tikz/every even column/.append style={column sep=0.4cm},
        legend style={at={(0,0)},anchor=center,draw=none,
          legend cell align={left},cells={line width=0.75pt}},
        legend image post style={sharp plot},
        legend cell align={left},
        ]
        \addplot [mark=triangle, OliveGreen!70] (0,0);
        \addplot [mark=o, Fuchsia!70] (0,0);
        \legend{bfloat16, sli-2.12};
      \end{axis}
    \end{tikzpicture}
  \end{center}
  \caption{Relative accuracy of bfloat16 and a 16-bit level-index representation compared with binary64.}
  \label{fig:representation-accuracy-bfloat16}
\end{figure}
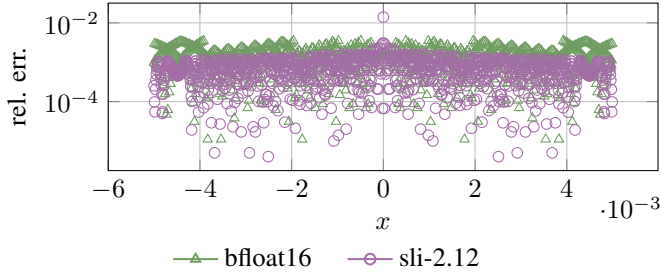

Figures~\ref{fig:representation-accuracy-binary16}~and~\ref{fig:representation-accuracy-bfloat16} show the accuracy of a 16-bit SLI arithmetic sli-2.12 compared with the binary16 and bfloat16 floating-point representations, respectively.
The accuracy was measured by comparing with binary64 in a narrow range of numbers around zero with a step size between the adjacent input samples of $10^{-5}$, computing the relative error.
The step size was chosen so that it is small enough to capture many values but big enough to visualize the errors in a plot.

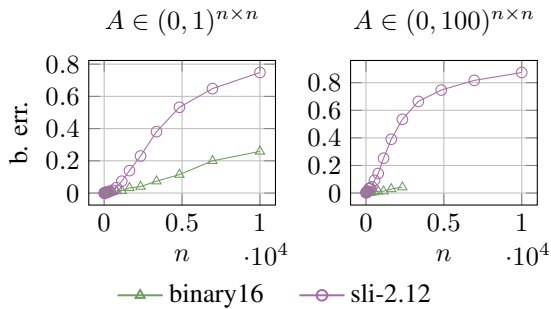
\begin{figure}[ht!]
  \begin{center}
    \begin{tikzpicture}
      \begin{groupplot}[
        group style={
          group size=2 by 1,
        },
        width=1.6in,
        grid=major,
        every x tick scale label/.style={
                  at={(axis description cs:0.95,-0.2)},
          anchor=north
        }
        ]
        
        \nextgroupplot[
        ylabel={b. err.},
        title={$A \in (0, 1)^{n\times n}$},
        xlabel = {$n$}
        ]
        
        \addplot[color=OliveGreen!70, mark=triangle] table [x=n, y=binary16] {data/matvec_binary16_0.dat};
        \addplot[color=Fuchsia!70, mark=o] table [x=n, y=level-index] {data/matvec_binary16_0.dat};

        \nextgroupplot[
        title={$A \in (0, 100)^{n\times n}$},
        xlabel = {$n$}
        ]
        
        \addplot[color=OliveGreen!70, mark=triangle] table [x=n, y=binary16] {data/matvec_binary16_1.dat};
        \addplot[color=Fuchsia!70, mark=o] table [x=n, y=level-index] {data/matvec_binary16_1.dat};
        
      \end{groupplot}
    \end{tikzpicture}

    \begin{tikzpicture}[trim axis left, trim axis right]
      \begin{axis}[
        title = {},
        legend columns=2,
        scale only axis,
        width=1mm,
        hide axis,
        /tikz/every even column/.append style={column sep=0.4cm},
        legend style={at={(0,0)},anchor=center,draw=none,
          legend cell align={left},cells={line width=0.75pt}},
        legend image post style={sharp plot},
        legend cell align={left},
        ]
        \addplot [mark=triangle, OliveGreen!70] (0,0);
        \addplot [mark=o, Fuchsia!70] (0,0);
        \legend{binary16, sli-2.12};
      \end{axis}
    \end{tikzpicture}
  \end{center}
  \caption{Backward error in $Ax$ with binary16 and sli compared with binary64.}
  \label{fig:matvec-accuracy0}
\end{figure}

\begin{figure}[ht!]
  \begin{center}
    \begin{tikzpicture}
      \begin{groupplot}[
        group style={
          group size=2 by 1,
        },
        width=1.6in,
        grid=major,
        every x tick scale label/.style={
          at={(axis description cs:0.95,-0.2)},
          anchor=north
        }
        ]
        
        \nextgroupplot[
        ylabel={b. err.},
        title={$A \in (0, 1)^{n\times n}$},
        xlabel = {$n$}
        ]
        
        \addplot[color=OliveGreen!70, mark=triangle] table [x=n, y=bfloat16] {data/matvec_bfloat16_0.dat};
        \addplot[color=Fuchsia!70, mark=o] table [x=n, y=level-index] {data/matvec_bfloat16_0.dat};

        \nextgroupplot[
        title={$A \in (0, 100)^{n\times n}$},
        xlabel = {$n$},
        ]
        
        \addplot[color=OliveGreen!70, mark=triangle] table [x=n, y=bfloat16] {data/matvec_bfloat16_1.dat};
        \addplot[color=Fuchsia!70, mark=o] table [x=n, y=level-index] {data/matvec_bfloat16_1.dat};
        
      \end{groupplot}
    \end{tikzpicture}
    
    \begin{tikzpicture}[trim axis left, trim axis right]
      \begin{axis}[
        title = {},
        legend columns=2,
        scale only axis,
        width=1mm,
        hide axis,
        /tikz/every even column/.append style={column sep=0.4cm},
        legend style={at={(0,0)},anchor=center,draw=none,
          legend cell align={left},cells={line width=0.75pt}},
        legend image post style={sharp plot},
        legend cell align={left},
        ]
        \addplot [mark=triangle, OliveGreen!70] (0,0);
        \addplot [mark=o, Fuchsia!70] (0,0);
        \legend{bfloat16, sli-2.12};
      \end{axis}
    \end{tikzpicture}
  \end{center}
  \caption{Backward error in $Ax$ with bfloat16 and sli compared with binary64.}
  \label{fig:matvec-accuracy1}
\end{figure}
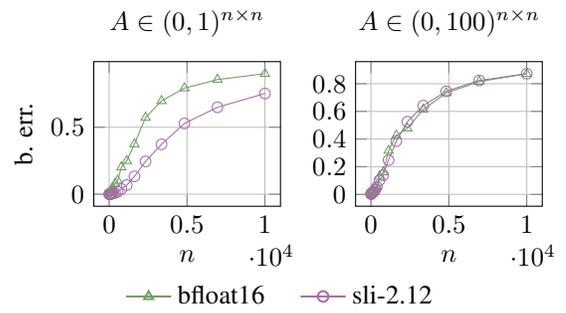

Figure~\ref{fig:matvec-accuracy0}~and~\ref{fig:matvec-accuracy1} show the relative backward error for matrix-vector multiplication $Ax$ with $A$ drawn from the two distributions shown and $x \in (0,1)^n$ for $n=[10, 10^4]$.
Binary16 demonstrates higher accuracy, but it overflows when $A\in(0,100)^{n\times n}$ whilst sli-2.12 continues computing.
On the other hand bfloat16 does not overflow in this particular experiment; sli-2.12 has better or equivalent accuracy compared with bfloat16.

\section{Conclusion}

SLI arithmetic simulator is presented which enables the community to experimentally study this number system.
In v0.1 we implemented most of the operators\footnote{\url{https://uk.mathworks.com/help/matlab/matlab_oop/implementing-operators-for-your-class.html}} for the \texttt{sli} objects.
Operators \texttt{mrdivide}, \texttt{mldivide}, \texttt{power}, \texttt{mpower}, \texttt{and}, \texttt{or}, and \texttt{not} are not yet implemented; the toolbox does not at present fully work in Octave. We plan extensions in the future versions of the toolbox.
Our goal is for the toolbox to act as an easy method for testing the accuracy of modern algorithms in SLI arithmetic which in turn may drive hardware architects to have another look at its implementation.

\section{Acknowledgements}

The author is grateful to the late N.~J.~Higham for discussions about the level-index arithmetic and M. Fasi for comments on the draft of this paper.
This work was supported by the EPSRC grant EP/P020720/1, and by the ECP (17-SC-20-SC), a collaborative effort of the U.S. Department of Energy Office of Science and the National Nuclear Security Administration.
The author also acknowledges the support of the School of Computing, University of Leeds.





\bibliography{references}
\bibliographystyle{siam}

\end{document}